\documentclass[sigconf]{acmart}
\usepackage{listings}
\usepackage{xcolor}
\usepackage{subcaption}
\usepackage{multirow}
\usepackage{enumitem}
\usepackage[export]{adjustbox}

\definecolor{codegreen}{rgb}{0, 0.6, 0}
\definecolor{codegray}{rgb}{0.5, 0.5, 0.5}
\definecolor{codepurple}{rgb}{0.58, 0, 0.82}
\definecolor{backcolour}{rgb}{0.95, 0.95, 0.92}

\lstdefinestyle{mystyle}{
    commentstyle=\color{codegreen}, 
    keywordstyle=\color{magenta}, 
    numberstyle=\tiny\color{codegray}, 
    numbersep=2pt,
    stringstyle=\color{codepurple}, 
    basicstyle=\ttfamily\footnotesize, 
    breakatwhitespace=false,         
    breaklines=true,                 
    captionpos=b,                    
    keepspaces=true,                 
    numbers=left,                    
    numbersep=5pt,                  
    showspaces=false,                
    showstringspaces=false, 
    showtabs=false,                  
    tabsize=2,
    frame=lines,
}

\setlist[itemize]{leftmargin=*}
\newcommand\recsysavailabilityurl{https://github.com/slow-steppers/NeighborHash}

\AtBeginDocument{%
  }

\copyrightyear{2024}
\acmYear{2024}
\setcopyright{acmlicensed}
\acmConference[CIKM '24]{Proceedings of the 33rd ACM International Conference on Information and Knowledge Management}{October 21--25, 2024}{Boise, ID, USA}
\acmBooktitle{Proceedings of the 33rd ACM International Conference on Information and Knowledge Management (CIKM '24), October 21--25, 2024, Boise, ID, USA}
\acmDOI{10.1145/3627673.3680034}
\acmISBN{979-8-4007-0436-9/24/10}

\begin{document}
\title{An Enhanced Batch Query Architecture in Real-time Recommendation}

\author{Qiang Zhang}
\orcid{0009-0005-7590-3567}
\affiliation{%
  \institution{Xi'an Jiaotong University}
  \city{Xi'an}
  \state{Shaanxi}
  \country{China}
}
\email{zhangqiang@stu.xjtu.edu.cn}

\author{Zhipeng Teng}
\orcid{0009-0000-5175-6762}
\affiliation{%
  \institution{Bilibili}
  \city{Shanghai}
  \country{China}
}
\email{tengzhipeng@bilibili.com}

\author{Disheng Wu}
\orcid{0009-0008-8278-0093}
\affiliation{%
  \institution{Bilibili}
  \city{Shanghai}
  \country{China}
}
\email{wudisheng@bilibili.com}

\author{Jiayin Wang}
\orcid{0000-0002-3862-6557}
\affiliation{%
  \institution{Xi'an Jiaotong University}
  \city{Xi'an}
  \state{Shaanxi}
  \country{China}
}
\email{wangjiayin@mail.xjtu.edu.cn}


\begin{abstract}
In industrial recommendation systems on websites and apps, it is essential to  recall and predict top-n results relevant to user interests from a content pool of billions within milliseconds. To cope with continuous data growth and improve real-time recommendation performance, we have designed and implemented a high-performance batch query architecture for real-time recommendation systems. Our contributions include optimizing hash structures with a cacheline-aware probing method to enhance coalesced hashing, as well as the implementation of a hybrid storage key-value service built upon it. Our experiments indicate this approach significantly surpasses conventional hash tables in batch query throughput, achieving up to 90\% of the query throughput of random memory access when incorporating parallel optimization. The support for NVMe, integrating two-tier storage for hot and cold data, notably reduces resource consumption. Additionally, the system facilitates dynamic updates, automated sharding of attributes and feature embedding tables, and introduces innovative protocols for consistency in batch queries, thereby enhancing the effectiveness of real-time incremental learning updates. This architecture has been deployed and in use in the bilibili recommendation system for over a year, a video content community with hundreds of millions of users, supporting 10x increase in model computation with minimal resource growth, improving outcomes while preserving the system's real-time performance.

\end{abstract}



\maketitle

\section{Introduction}

\begin{figure}[h]
    \centering
    \includegraphics[width=1\linewidth]{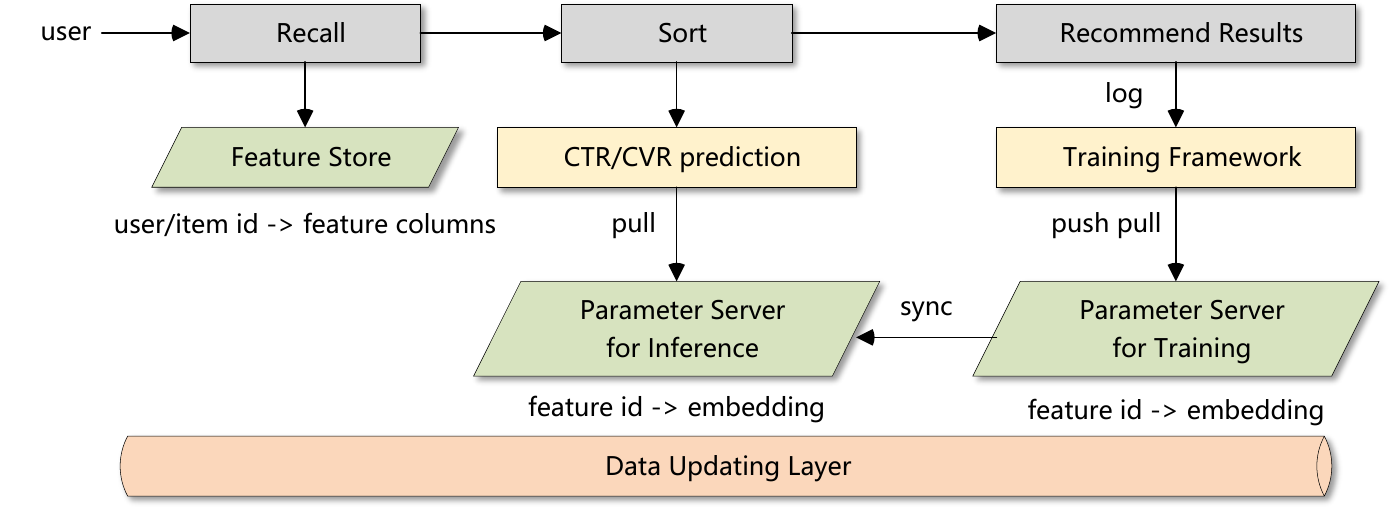}
    \caption{Key-Value Storage in Recommendation system}
    \label{fig:kv-in-recsys}
\end{figure}

In recommendation systems, the rapid computation of item relevance to users—ideally within milliseconds is crucial for optimal performance, and storage systems capable of managing vast data volumes and facilitating high-speed batch queries are one of the most vital components, as illustrated in \autoref{fig:kv-in-recsys}. Whether in the recall or ranking phases, or during offline model training~\cite{li2014scaling,li2013parameter}, more features and faster computation generally lead to better recommendation outcomes. Redis~\cite{RefWorks:RefID:15-redis} is noted for its straightforward architecture and robust efficiency, proving effective in supporting recommendation systems. However, in larger-scale industrial settings, developing storage engines that enhance rapid batch query capabilities for recommendations remains a critical research area.

As deep learning becomes the mainstay in the realm of recommendation systems~\cite{covington2016deep, cheng2016wide, gupta2020architectural,zhao2019recommending,davidson2010youtube}, the size of these models has escalated, now reaching terabyte scale~\cite{lian2022persia}. Deep learning-based recommendation systems require a vast array of features to accurately depict complex user behaviors, attributes, and preferences. Bilibili primarily distributes video content, typically around 10 minutes in length. Therefore, compared to e-commerce recommendations, features such as key frame image embeddings, video sentiment, and the author's attitudes and opinions are crucial for content delivery. 

In industrial recommendation systems with hundreds of millions of users and billions of items, each record may comprise hundreds of simple floating-point features, potentially leading to embedding table sizes in the terabyte range. Such extensive feature storage and embedding tables cannot be housed within a single memory unit. This leads to the conclusion that optimizing the batch query service of recommendation systems necessitates addressing three key challenges.

\begin{itemize}
    \item \textbf{High Performance.} A storage engine with high-speed batch reading capabilities serves as the carrier for a vast array of feature values and embedding tables, enhancing the throughput during model training and prediction.
    \item \textbf{Scalability.} A distributed query services cluster maintains high throughput even as system size increases - both in terms of the number of features and the volume of items processed. 
    \item \textbf{Flexibility and Consistency.} Recent user behaviors with session-segment are more effective in predicting future actions~\cite{tuan20173d,lu2018like,twardowski2016modelling,Shen_2022}. rapid updates ensure that real-time recommendation systems maintain effectiveness by reflecting users' current interests. It is essential to avoid data inconsistencies during updates to prevent losses in recommendation system performance.
\end{itemize}

In this paper, we design and implement a distributed query architecture for batch query that significantly enhances recommendation system performance. Here is our contribution. First, we have made several design to improve Coalesced Hashing, resulting in a new hash table structure called NeighborHash, which minimize the number of cacheline accesses per query to achieve larger query throughput within the constraints of limited memory bandwidth. A comprehensive performance evaluation of NeighborHash reveals that, in batch query scenarios, it achieves a 170\% performance improvement compared to the most optimized hash-tables, Validation experiments confirm the effectiveness of our optimizations. An open-source version of neighborhash is available at the following address \url{\recsysavailabilityurl}. Second, Building on Neighborhash, we have introduced a SSD-based distributed key-value storage service that supports horizontal scaling of features and model sizes while maintaining low resource consumption. Particularly focused on the multi-version states of data during real-time learning updates, we optimized update and query protocols to ensure strong data consistency during batch queries, thereby maintaining stable recommendation performance. Finally, we deployed this architecture in the bilibili recommendation system and achieved significant benefits.

\section{System Design}

\begin{figure}[h]
    \centering
    \includegraphics[width=1\linewidth]{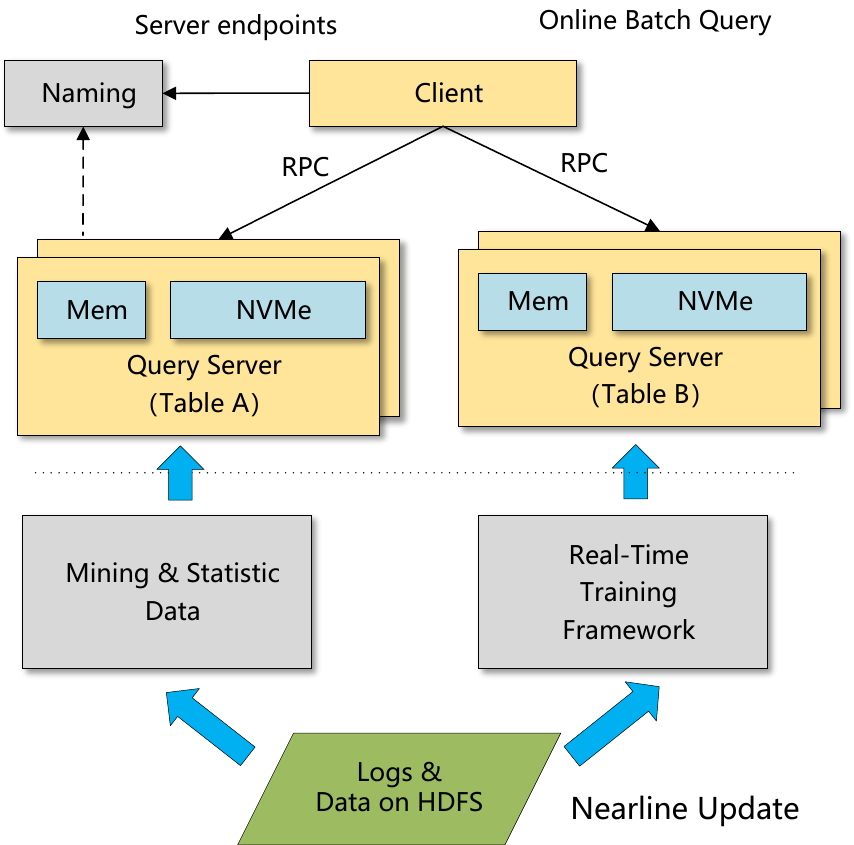}
    \caption{Batch Query Architecture for Multi-Data Lookup}
    \label{fig:key-value-cluster-arch}
\end{figure}

The design of the serving architecture draws significant inspiration from Google Mesa~\cite{RefWorks:RefID:32-gupta2014mesa:}. The system comprises the Batch Query Subsystem and Update Subsystem, outlined in \autoref{fig:key-value-cluster-arch}. As a typical architecture in recommendation systems, the Update Subsystem manages data updates, encompassing user behavior properties and parameter updates for recommendation model training. The Query Subsystem caters to extensive data volumes and high concurrency requirements through a distributed layout with multiple shards and replicas. This setup is designed to handle a significant request volume (peaking at approximately 100k qps) and enormous feature tables that cannot be deployed on a single machine. At its foundation lies our hash table, Neighborhash, optimized for batch querying in recommendation systems and described below.

\subsection{NeighborHash}

\subsubsection{Hashtable structure}
NeighborHash, based on Coalesced Hashing, utilizes a flat array to store all buckets, with each bucket containing key, value and the index to the next spot in the chain or else the null value. To minimize the number of cachelines involved in queries, we implemented the following design:

\textbf{Lodger relocation}. In traditional Coalesced hashing, hash table buckets are numbered 1 to M'. The first M buckets serve as the hash function's address region, while the remaining M'–M buckets are exclusively for colliding records, known as the cellar. When the cellar fills up, subsequent colliders must occupy empty buckets in the address region, potentially leading to further collisions with later-inserted records.  Hence, the PSL (probing sequence length) is sensitive to the cellar region's size. To minimize the PSL, we eliminate the cellar region allocation. Instead, we place conflicting elements in the address region and dynamically adjust them. The method is as follows: For a record \textit{x} in bucket \textit{i}, if Hash(x.key) is \textit{i}, it is termed as the \textit{host} record; otherwise, it is termed as the \textit{lodger} record. When inserting a new record \textit{y}, with Hash(y.key) resulting in bucket \textit{j}, if bucket \textit{j} is occupied by a lodger, a vacant position is sought to relocate the lodger before storing record \textit{y} in bucket \textit{j}. If bucket \textit{j} is occupied by a host, a vacant position is sought to store record \textit{y}, and it is appended to the end of the chain.

\begin{figure}[h]
    \centering
    \includegraphics[width=0.9\linewidth]{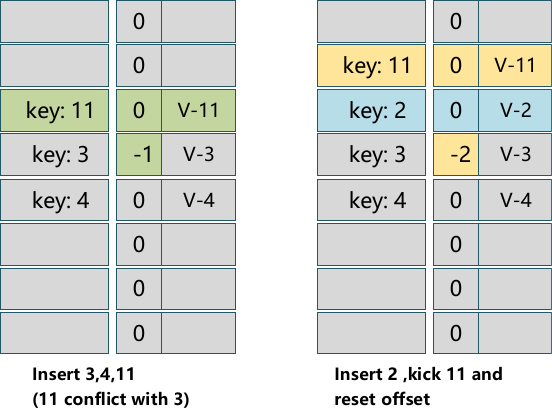}
    \caption{Insertion of NeighborHash with Lodger Relocation}
    \label{fig:hash-insert-kickout}
\end{figure}

Lodger relocation can be seen as a dynamic cellar strategy, ensuring minimization of PSL (same as separate chaining). However, the drawback is that the insertion process becomes more complex. Considering query requests dominate the workload of recommendation systems, we believe this trade-off is worthwhile. The aforementioned process can be referenced in \autoref{fig:hash-insert-kickout}.

\begin{figure}[h]
    \centering
    \includegraphics[width=1\linewidth]{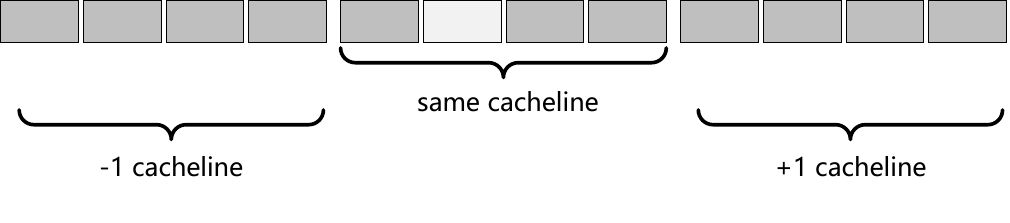}
    \caption{Find available node in neighbor cacheline}
    \label{fig:neighbor-cacheline}
\end{figure}

\textbf{Cacheline-aware neighbor probing}, NeighborHash strives to place buckets on the same chain within the same cache line to minimize memory bandwidth usage for each query. During the search for available buckets, the algorithm first examines buckets within the same cacheline. Unlike Linear Probing, which probes only in one direction, NeighborProbing conducts bidirectional probing within the cacheline. If none is found, the search expands bidirectionally to identify the nearest available bucket to the head, as illustrate in \autoref{fig:neighbor-cacheline} . This approach minimizes cross-cache-line probes during queries. Experimental results demonstrate that, on a random dataset of 100 million entries with a load factor of 75\%, the average number of cache line accesses per query is approximately 1.12. Compared to Linear Probing, which is 1.47, Cacheline-aware Neighbor probing requires fewer cachelines. If no available bucket is within the same cacheline, relocating to nearby cachelines may mitigate TLB cache misses and lead to a smaller relative offset to the previous record in the chain, enabling offset compression.

\begin{figure}[h]
  \centering
  \includegraphics[width=1\linewidth]{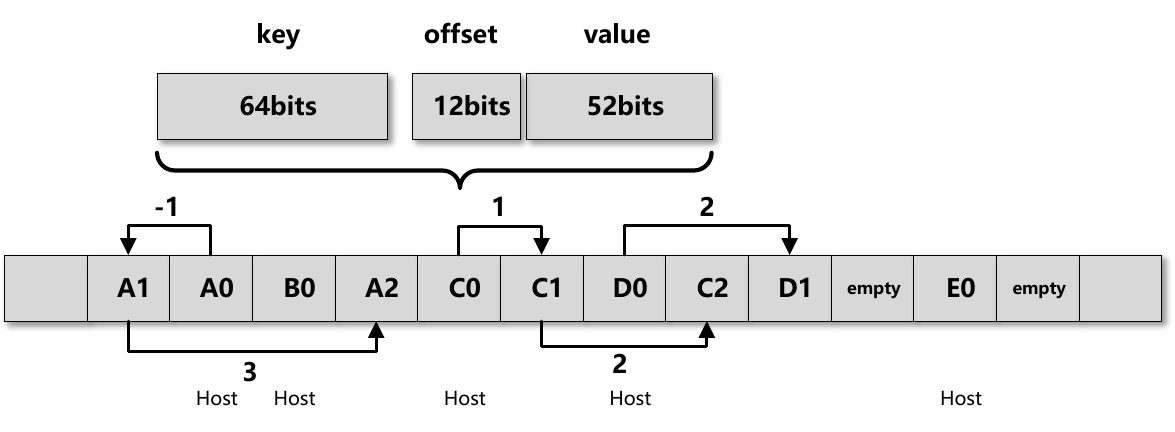}
  \caption{Data Structure}
  \label{fig:hash-data-structure}
\end{figure}

\textbf{Inline chaining}, We utilize the high 12 bits of NeighborHash's value field to represent relative offsets, enabling the implementation of a conflict linked list. The range of representation is -2047 to 2048. Considering the conflict allocation strategy primarily aims to find a suitable location near the root node, the 12-bit relative offset is more than sufficient. In practice, a 12-bit offset can achieve a load factor of over 80\%. Consequently, NeighborHash employs a storage representation of 52 bits for the actual value, which is typically used to store pointers or offset values in recommendation system, illustrated in \autoref{fig:hash-data-structure}.

\textbf{Lookup Acceleration}.
Hash-table lookup is a predominant operation in recommender systems. Vectorization~\cite{RefWorks:RefID:23-polychroniou2015rethinking} and AMAC~\cite{RefWorks:RefID:22-kocberber2015asynchronous} have been extensively utilized for optimizing Hash-table batch lookup. We implemented inter-query vectorized query on NeighborHash using the IMV(Interleaved Multi-Vector)~\cite{RefWorks:RefID:35-fang2019interleaved} method. Efficient SIMD compress and expand operations can be employed for the append and fill processes. Vectorization can bring significant throughput improvements on small datasets, but on larger datasets, memory latency becomes the predominant factor. We implemented the AMAC method on NeighborHash and conducted experiments. The experiments show that on large datasets, with the help of AMAC, NeighborHash can achieve almost a doubling of throughput compared to its original performance. 

\subsubsection{NVMe storage}
\begin{figure}[h]
    \centering
    \includegraphics[width=1\linewidth]{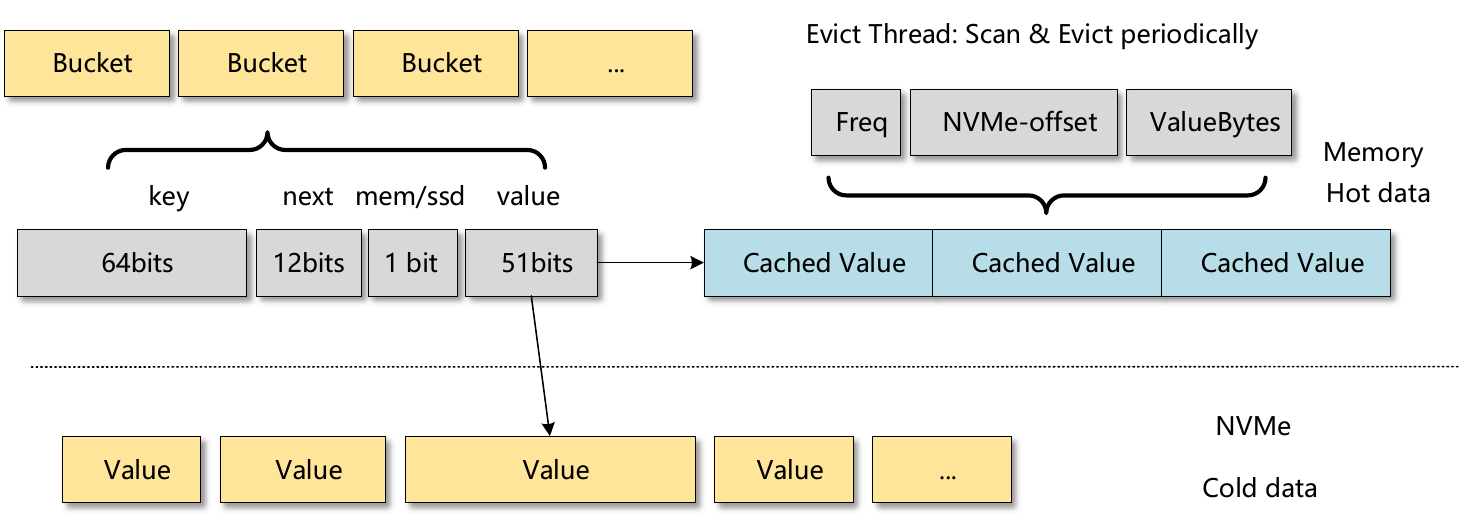}
    \caption{NeighborHash structure with NVMe support}
    \label{fig:nhash-nvme-supp}
\end{figure}

In industrial scenario with billions of users and items, there exists a vast amount of cold data. Compared to storing all data in memory, it is cost-effective to store cold data in NVMe and hot data in memory~\cite{wan2021flashembedding}. To maintain the latency of request responses, we store index(key to value offset) in memory using the NeighborHash structure and store the value bytes in a two-tier manner. The 52-bit payload in NeighborHash includes 1 bit to indicate whether the data is stored in memory or NVMe. The payload of hot data points to the memory containing the LRU metadata, while that of cold data points to the file system offset. Eviction is completed by an asynchronous thread scanning the metadata of hot data. As illustrated in \autoref{fig:nhash-nvme-supp}. Storing both hot and cold keys in memory reduces the concurrent read/write overhead on the Hash-table associated with traditional LRU methods. Additionally, during a cache miss, typically only one or a few NVMe I/O operations are involved. We believe that the additional memory overhead is more economical compared to the savings in CPU usage and the improvement in throughput. Due to space limitations, we will not provide a detailed analysis here.

\subsection{Batch Query Subsystem}

The Batch Query Subsystem is architected as a multi-sharded, multi-replica framework. Given the unique update cycles, query loads and geographic distribution demands of each table in the business domain, the subsystem is organized based on individual tables. Each table maps to a distinct query service, with specific shard and replica information maintained by a metadata service. This underlying infrastructure is supported by etcd~\cite{etcd20} for robustness.

\subsubsection{Automatic Sharding}

Sharding data during service has two primary advantages:
\begin{enumerate}
    \item It avoids the challenges linked to overly large service instances, including lengthy start-up times and high data retrieval bandwidth, enhancing system stability. Smaller shards also facilitate quicker migration and recovery.
    \item Parallel queries across multiple shards reduce the batch query load on individual instances. Coupled with asynchronous processing for immediate tasks like click-through rate estimation, this approach notably decreases user request latency.
\end{enumerate}
The system automatically manages shard creation based on set configuration parameters, ensuring no shard exceeds its designated size. Should the table grow or shrink during updates, re-sharding occurs during the next update cycle, with updated metadata synchronized across the live cluster.

\subsubsection{Rolling Update and Query Consistency}

\begin{figure*}[h]
    \centering
    \includegraphics[width=1\linewidth]{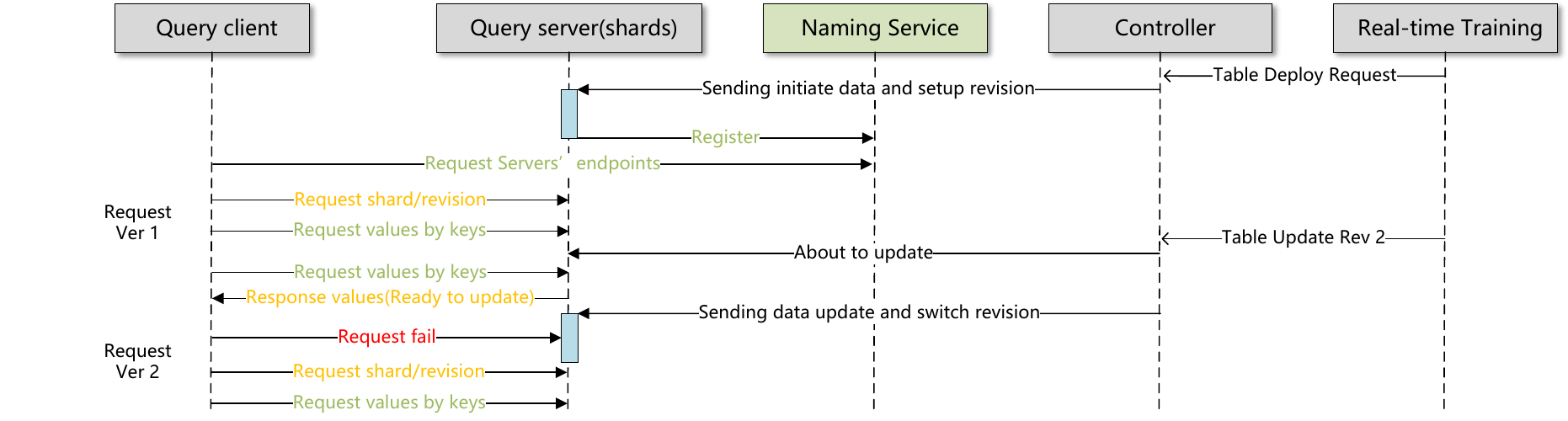}
    \caption{Data Online Updating and Query Interaction}
    \label{fig:kfc-req-update}
\end{figure*}

Real-time recommendation systems in production cannot have downtime for updates and maintenance, as they need to be constantly operational. Typically, updating data requires additional resources. A straightforward approach involves deploying a backup query service for each table to handling all traffic, switching to it once ready, necessitating double the resources. We implement a rolling update approach, updating one replica at a time and only requiring an additional 1/n of the resources. However, this introduces the issue of consistency across different versions over shards during table updating. For some model embedding tables, only values from the same training batch are comparable, which we refer to as strong version data. For sorting, it is crucial that features for the same batch of items come from the same version to ensure comparability.

\begin{figure}[h]
    \centering
    \includegraphics[width=1\linewidth]{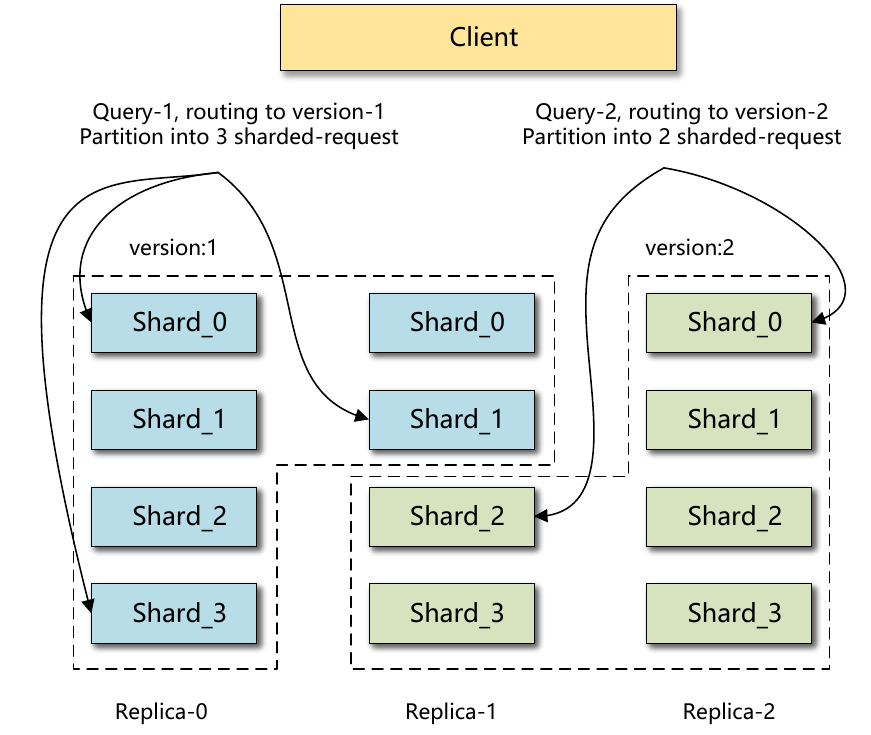}
    \caption{Query with version control during updating}
    \label{fig:ver-client-req}
\end{figure}

Typically, shard and data version information stored on servers are registered with naming service and updated regularly. Clients consult this naming service for details, and request version-specific data from corresponding servers. However, in industrial-scale distributed systems with thousands of client and server instances, network delays and packet losses can prolong the process of server metadata updating in the Naming service and its retrieval by clients. This delay prevents servers from providing immediate service after they are ready, requiring explicit version consistency confirmation between clients and servers before proceeding with the next batch of rolling updates. This significantly extends the total update time, especially for data services that require frequent updates, where clients may not yet detect the previous version before the server needs to load new data, leading to request failures due to version inconsistencies. To address this issue, we utilize the Naming service only for updates to the server instance interfaces (IP:port), such as additions or deletions, while shard and version metadata are communicated directly between clients and servers through the query protocol, ensuring strong consistency in shard management on the client-side. The data update mechanism and access consistency scheme are illustrated in \autoref{fig:kfc-req-update} and \autoref{fig:ver-client-req}, respectively. Data compression and asynchronous pipeline processing are also encapsulated within this library. Moreover, it provides direct access to embedding tables and feature storage, minimizing network bandwidth overhead compared to a proxy-based approach.

\section{Experiment and Result}

In this section, we first provide a detailed evaluation and comparison of the optimization of hashtable, as this is a key factor in improving query throughput capacity. The second part directly assesses the latency optimization and resource efficiency from the new query cluster by conducting online comparisons in the recommendation system of bilibili, along with evaluating the impact of data consistency on performance.

\subsection{Evaluation of NeighborHash}
Firstly, we designed comparative experiments between NeighborHash and existing scalar hashmaps. The focus was on conducting ablation experiments and analysis between NeighborHash and Coalesced Hashing. Subsequently, we conducted comparisons and analyses of different vectorization methods and implementations.

\subsubsection{Experiments Setup}
We conducted all experiments on a Linux Server, which is configured as outlined in README of github repo, The live recommendation system is also deployed on machines of these specifications. In the context of recommendation systems, the specific type of hash table we focus on that includes skewed data distribution, large dataset size, high load factor, very high read/write skew, and a high successful lookup ratio. Certain configurations can influence the evaluation results, as detailed below.

\begin{itemize}[leftmargin=*]
    \item Hardware prefetch, In certain scenarios, for precise analysis the efficiency of hash-table, we disabled hardware prefetching.
    \item Hash function, the hash function employed across all tests was absl::Hash<uint64\_t>. 
    \item Size and Load Factor, as mentioned earlier, we tested hash-tables with keys and values both being 64 bits, thus each element occupying 16 bytes of memory. We opted for datasets of sizes 256KB(16K), 2MB(128K), 16MB(1M), 256MB(16M), 2GB(128M) and 16GB(1G) for testing. We ensure that the hash-table's load factor is set to 80\%, a practical value.
\end{itemize}

Data distribution significantly impacts the test results of the Hash-table. Due to space limitations and the typically high memory load in real-world scenarios, our tests and analyses were conducted on uniformly distributed datasets. We also tested skewed data, and the conclusions were consistent with those from the uniformly distributed data in terms of trends and qualitative analysis.

\textbf{Metrics}. We measure hash-table throughput in MOPS (Million Operations Per Second) and monitor memory bandwidth consumption in BPL (Bytes Per Lookup). On Intel platforms, we utilize PCM to track bandwidth and calculate BPL by dividing bandwidth (BPS - Bytes Per Second) by the MOPS. Our evaluation also considers the LLC (Last Level Cache) cache-miss rate. We analyze probing performance using the \textit{Average Probing Cache Lines (APCL)} metric, indicating the average number of cache lines accessed for each successfully found key. For exploring the upper limit of hash-table lookup operations, we created a hash-table with zero collisions, albeit without guaranteed correctness. Each lookup involved hash calculation and random reading, deemed indispensable minimal operations in hash-table lookup. This imperfect hash-table version is referred to as RA (Random Access). Subsequent experiments showed that NeighborHash achieved 90\% of RA's throughput with parallel optimization.

\subsubsection{Scalar Hash-Tables Comparsion}
we evaluate the scalar hash-tables. The hash-tables evaluated include Linear Probing, Coalesced hashing, ska::bytell\_hash\_map~\cite{RefWorks:RefID:38-2018a} and Neighborhash. We also introduce the previously mentioned random access in the comparison to understand the absolute level of query performance. The results are shown in \autoref{tab:scalar-hash}. Due to the significant throughput of RA on datasets smaller than 256MB, it has been omitted from the table. As the dataset size increases, continuous cache misses lead to a sustained decline in throughput. In comparison to other implementations, NeighborHash consistently exhibits performance improvements across all scenarios. At a dataset size of 16GB, NeighborHash demonstrates over 50\% higher query throughput compared to other implementations. In this test, the success query rate(SQR) was 90\%, consistent with mainstream recommendation systems online. We also conducted tests under conditions of low hit rates(30\%), and the results showed completely consistent trends.

\begin{table}[h]
\centering
\begin{tabular}{llllll}
\toprule
\multirow{2}{*}{Hashtable} & \multicolumn{5}{l}{Dataset size} \\
 & 256KB & 2MB & 16MB & 256MB & 16GB \\
\midrule
Linear probing & 38 & 29 & 22 & 12 & 11 \\
ska::bytell\_hash\_map & 72 & 53 & 38 & 20 & 19 \\
Coalesced hashing & 92 & 56 & 48 & 21 & 19 \\
\textbf{Neighborhash} & \textbf{116} & \textbf{74} & \textbf{66} & \textbf{37} & \textbf{36} \\
Random Access / & / & / & / & 67 & 67  \\
\bottomrule
\end{tabular}
\caption{Scalar Hashtable Lookup Performance(Mops)}
\label{tab:scalar-hash}
\end{table}

To analyze the impact of dataset size on hash-table performance, we conducted a detailed evaluation of NeighborHash, as presented in \autoref{tab:nhash-eval-uniform}. For dataset sizes less than 2MB, most memory loads can be accommodated by the L1 and L2 cache. At this point, the MOPS is more influenced by the number of instructions and their cycles. Starting from a size of 16MB, the LLC miss rate begins to rise, reaching about 34\% at 32MB. Simultaneously, MOPS starts to decline, while Bytes-per-lookup increases. As the LLC miss rate increases, Bytes-per-lookup exhibits a rapid growth after 32MB (L2 cache size) and converges at a dataset size of 2GB, indicating that MOPS is almost entirely dominated by memory latency. 

\begin{table}[h]
    \centering
    \begin{tabular}{ccccc}
        \toprule
        Datasets & LLC-LD & LLC-MR & MOPS & BPL\\
        \midrule 
         256KB & 0.004 & 25\% & 116 & 0.15\\
         2MB & 54 & 0.01\% & 74 & 0.15\\
         16MB & 82 & 3.58\% & 66 & 2.56\\
         32MB & 59 & 33.98\% & 48 & 27.9\\
         256MB & 46.6 & 90.9\% & 37 & 78.7\\
         2GB & 45.5 & 98.6\% & 37 & 81.4\\
         16GB & 44.3 & 99.2\% & 39 & 82.1\\
         \bottomrule
    \end{tabular}
    \caption{NeighborHash, SQR=90\%}
    \label{tab:nhash-eval-uniform}
\end{table}

\textbf{Ablation Analysis}.To comprehend the impact of various design components of NeighborHash on outcomes, we conducted ablation analysis on the following three key designs of NeighborHash:

\begin{itemize}[leftmargin=*]
    \item Lodger relocation, employing only this strategy in Coalesced hashing, the search efficiency is equivalent to Coalesced hashing with perfect cellar, abbreviated as PerfectCellarHash. 
    \item Cacheline-aware neighbor probing, building upon PerfectCellarHash, prioritizes searching for available buckets near the cacheline of the last node in the conflict chain and its vicinity. Relative offsets are stored in a separate offset array. This implementation is referred to as NeighborProbing.
    \item Inline-chaining, based on NeighborProbing, encodes relative offsets into the high 12 bits of the value, thus completing the full implementation of NeighborHash.
\end{itemize}

We conducted experiments and analyses on different datasets using the three aforementioned implementations. Datasets larger than 2GB exhibited similar trends. As we are particularly interested in the performance on larger datasets, the \autoref{tab:nhash-ablation-large} presents the results for the 16GB dataset.

\begin{table}[h]
    \centering
    \begin{tabular}{ccccc}
        \toprule
        Hashmap & MOPS & MOPS-gain & APCL\\
        \midrule 
         CoalescedHashing & 19 & 1 & 1.72 \\
         PerfectCellarHash & 23 & 1.21 & 1.48 \\
         NeighborProbing & 30 & 1.30 & 1.34 \\
         NeighborHash & 39 & 1.30 & 1.14\\
         \bottomrule
    \end{tabular}
    \caption{size=16GB, SQR=90\%, LF=0.8, Ablation Analysis}
    \label{tab:nhash-ablation-large}
\end{table}

Experimental results indicate that the aforementioned three key designs of NeighborHash contributed to throughput gains of 20\%, 30\%, and 30\%, respectively, in terms of Millions of Operations Per Second (MOPS). It is noteworthy that these three designs are not entirely independent; the joint action of lodger relocation and neighbor probing enables offset compression, which, when combined with specific usage scenarios, allows for its integration into the value. Average Probing Cachelines(APCL) decreased from 1.72 to 1.14, resulting in saved memory bandwidth and achieving the goal of throughput improvement. We also evaluated the APCL of linear probing with Lodger Relocation, resulting in 1.24. It can be inferred that bidirectional probing can contribute approximately 9\% to memory bandwidth efficiency compared to unidirectional probing.

\begin{figure}[h]
    \centering
    \includegraphics[width=1\linewidth]{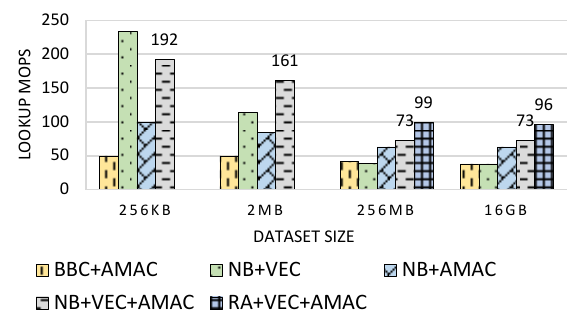}
    \caption{NB with Vec and/or AMAC,SQR=90\%}
    \label{fig:nbhash-vec-eval}
\end{figure}

\textbf{Integration with Optimization}. We implemented inter-query vectorization on NeighborHash, and the evaluation results are presented in \autoref{fig:nbhash-vec-eval} which also includes the evaluation results of AMAC combined with SIMD for NeighborHash. For queries on smaller datasets, SIMD optimizations are most effective because most of the data can be directly stored in L2-cache. As the dataset size increases, AMAC becomes the better choice. In practical system optimizations, selecting the appropriate version based on the size of the dataset can yield the best results.

\subsection{Benefits in Industrial Scenario}
We deployed the batch query architecture proposed in this paper in Bilibili recommendation system. The significant improvements in computational and storage capabilities have facilitated optimizations across the recommendation models. Here, we present two key online comparison metrics: access latency and impact of data consistency on effectiveness.

\textbf{Latency}. We compared its effectiveness with a RocksDB-based Key-value query service, a commonly used solution in industrial recommendations that our system has previously utilized. We observed a 3 to 6 times increase in query latency compared to the baseline. In an experiment focusing on a high-traffic storage table for item features(40M items, 1KB per-item), with a peak online Key-Seek Per Second (KPS) of around 700k, we found that as the batch\_size of single key queries increased, performance of query service remained stable without significant degradation, as illustrate in \autoref{tab:latency-kv}. CPU profiling with the perf tool showed that hash lookups in Neighborhash consumed a small portion of the CPU, with about half of the CPU usage dedicated to IO operations like payload packaging. Conversely, in the RocksDB implementation with an in-memory table configured as a hashtable and 10GB of memory(same with NeighborKV), approximately 30\% of CPU utilization was allocated to memory queries and retrieval. Consequently, as the batch size increased, the performance discrepancies became more noticeable.

\begin{table}[h]
    \centering
    \begin{tabular}{ccc}
        \toprule
         Key-value & Batch-size & AVG-latency(ms) \\
        \midrule 
         & 10 &  1.11 \\
         KV(Rocksdb) & 100 &  10.56 \\
         & 500 &  25.81 \\
        \midrule 
         & 10 &  1.05 \\
         KV(Neighborhash) & 100 &  1.78 \\
         & 500 &  3.31 \\
         \bottomrule
    \end{tabular}
    \caption{KV(Rocksdb) vs KV(Neighborhash)}
    \label{tab:latency-kv}
\end{table}

\begin{figure}[h]
    \centering
    \includegraphics[width=0.9\linewidth]{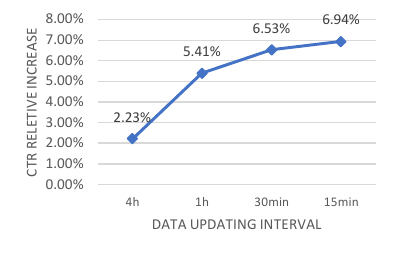}
    \caption{CTR Relative increase with consistency policy}
    \label{fig:ctr_inc_version}
\end{figure}

\textbf{CTR Improvements with data consistency}. In our A/B online experiments in the bilibili recommendation system, we compare click-through rate variances when enforcing consistent data version constraints during data shard replica rolling updates. The overall data comparison spans a full day, i.e., 24 hours, with results presented in \autoref{fig:ctr_inc_version}. It is evident that the shorter the update interval, the more pronounced the benefits of batch query data consistency, as more inconsistencies occur with frequent updates. We analyzed the scenario where multi-shard updates of the embedding table occur without a consistency protocol, leading to approximately 3\% inconsistency in results. This inconsistency implies that a single estimation might utilize multiple versions of the embedding weights. A detailed analysis of the inconsistent cases reveals that discrepancies among correlated features significantly impair the estimation results. Therefore, we conclude that the improvement in CTR can be attributed to the enhancement in consistency. While the specific effects may vary across systems and attribute tables, observations in various systems and multidimensional data showcase significant performance improvements through maintaining consistency.

\textbf{Resource Saving}. NVMe usage depends on data categorization into hot and cold data, which varies across applications. A feasible recommendation is to employ NVMe storage for embedding tables and features that exhibit large data volumes but with very low Key-Seek Per Second (KPS) rates. The resource advantages here mainly come from the improvements in high-performance batch queries. The boost in single-instance throughput allows for fewer replicas, reducing the number of redundant resources due to rolling update capability. we has saved around 30\% of machine resources in our recommendation system. 
In systems with a long-tail distribution of user activity and item popularity, certain low-traffic embedding tables still require two replicas for fault tolerance. Future optimizations could focus on cluster scheduling to improve resource efficiency by sharing replicas among low-traffic tables.

\section{Related Work}

Various evolutionary schemes have been developed based on hash strategies~\cite{RefWorks:RefID:9-pagh2004cuckoo,RefWorks:RefID:7-herlihy2008hopscotch,RefWorks:RefID:8-celis1985robin}. Coalesced Hashing~\cite{vitter1982implementations} is an attempt to combine linear probing and separate shaining but lacks consideration for cacheline friendliness and performs poorly in current practical environments. Neighborhash addresses these limitations and demonstrates significant improvements. Notable examples like Absl's flat\_hash\_map~\cite{abseilio2017abseilio} and Facebook's F14~\cite{RefWorks:RefID:34-bronson2019open-sourcing} have excelled in optimizing query efficiency through SIMD instruction utilization. Experimental evaluations show that NeighborHash achieves lower Average Probing Cache Lines and higher query throughput compared to linear probing. Monolith~\cite{liu2022monolith} effectively filters low-frequency and outdated feature IDs in recommendation systems, reducing conflicts and enhancing model performance by utilizing Cuckoo hashing, while ~\cite{zhang2020model} give a hybrid hashing method to combine frequency hashing and double hashing techniques for model size reduction.

Due to variations in scenarios and applications, there is no one-size-fits-all design for key-value storage. The flexibility of key-value storage leads to ongoing research in optimization tailored to different contexts. Implementations like LevelDB, RocksDB, FlashStore, and SILT~\cite{RefWorks:RefID:12-leveldb,rocksdb2021,dong2021rocksdb,RefWorks:RefID:30-debnath2010flashstore:,RefWorks:RefID:13-lim2011silt:} utilize LSM-trees for in-memory key-value storage. SlimDB~\cite{RefWorks:RefID:14-ren2017slimdb:} employs dynamic compaction and in-memory index optimizations to enhance throughput for semi-sorted data, while F2~\cite{kanellis2023f2} separates hot and cold data domains to boost overall performance under large data skew. In the realm of recommendation systems, feature store systems like Uber's Michelangelo Palette, Google Feast and Amazon SageMaker Feature Store~\cite{palette-meta-store-journey,google-Feast,aws-sagemaker} offer solutions for AI applications. RecShard~\cite{sethi2022recshard} shards embedding tables based on training data distribution and model characteristics to improve model training throughput. EVTable~\cite{kurniawan2023evstore} implemented a three-layer embedding lookup table to enhance recommendation effectiveness and optimize resources. These optimizations differ slightly from the approach in this paper, which enhances online inference throughput significantly within its feature store architecture by employing shard constraints and ensuring access consistency, suggesting that integrating the designs could yield greater benefits.

\section{Conclusion}

This paper presents an enhanced batch query architecture tailored for industrial-grade recommendation systems, providing high-performance throughput for batch queries of user and item features, model parameters and embedding tables.

We introduce NeighborHash, a hash-table optimized for batch point queries. By strategically placing conflicting nodes in neighboring positions, it reduces search probe times and cache misses, significantly enhancing batch query performance. Comparative analysis and ablation experiments demonstrate that Neighborhash outperforms existing hashtable structures significantly in recommendation scenarios. Building upon Neighborhash, we develop query cluster, which efficiently saves resources through NVMe storage compatibility and rolling updates. It ensures the real-time update of model parameters crucial for recommendation effectiveness while maintaining access consistency during updates. We deployed this system in the bilibili recommendation system and achieved a win-win in both resources and performance.


\bibliographystyle{ACM-Reference-Format}
\balance
\bibliography{nhash}

\end{document}